# Molybdenum oxide hole selective transport layer by hot wire oxidation-sublimation deposition for silicon heterojunction solar cells

Fengchao Li[1,2], Yurong Zhou[1]*, Ming Liu[1,2], Gangqiang Dong[1], Fengzhen Liu[1,2]*, Wenjing Wang[3], Donghong Yu[4]

1. College of Materials Science and Opto-Electronic Technology, University of Chinese Academy of Sciences, 100049, Beijing, China
2. Sino-Danish College, University of Chinese Academy of Sciences, 100190, Beijing, China
3. Institute of Electrical Engineering, Chinese Academy of Sciences, 100190, Beijing, China
4. Department of Chemistry and Bioscience, Aalborg University, DK-9220, Aalborg, Denmark

Abstract

In this article, a novel hot wire oxidation-sublimation deposition (HWOSD) technique was developed to prepare molybdenum oxide ($MoO_x$) thin films with high quality. Silicon heterojunction (SHJ) solar cells with the HWOSD $MoO_x$ as a hole selective transport layer (HSL) were fabricated. Thickness of the $MoO_x$ layer and annealing process of the solar cells were studied and optimized. A power conversion efficiency up to 21.10% was achieved on a SHJ solar cell using a 14nm $MoO_x$ layer as the HSL. Dark current density-voltage-temperature (*J-V-T*) characteristics of the SHJ solar cell were measured at the temperatures from 200K to 380K. Transport processes including thermionic emission of electrons over the potential barrier and quantum assisted tunneling of holes through the gap states in the $MoO_x$ layer were proposed for the $MoO_x$/n c-Si heterojunction. The investigation of the transport mechanisms provides us a better understanding of the characteristics of the novel SHJ solar cells and it is helpful for us to fully demonstrate the potential of such kind of solar cells in the future.

Keywords: Hot wire oxidation-sublimation deposition, Molybdenum oxide thin film, Hole selective transport layer, Silicon heterojunction solar cell, Transport mechanisms

**1. Introduction**

More and more attention has been paid to the novel silicon heterojunction (SHJ)

solar cells by making use of metal oxides to replace the conventional doped hydrogenated amorphous Si (a-Si:H) thin films as the carrier selective transport layers[1-5]. Due to the wide band-gap nature, optional work functions and relatively simple fabrication techniques related to the metal oxides, the novel SHJ solar cells show great potential to further improve the efficiency and reduce the cost of c-Si based solar cells[3,6]. Difference between work functions of the metal oxides and Fermi level of the c-Si leads to a large energy band bending of the c-Si near the interface, which allows only one type of carriers to pass through and inhibits carrier recombination at the interface[7]. Some metal oxides with high work functions, such as $MoO_x$[3,4], $VO_x$[8-10], $WO_x$[11,12], $NiO_x$[13,14] and $CuO_x$[15,16], can provide a good hole selective transport when they form heterocontacts with c-Si. Similarly, some metal oxides with low work functions, e.g. $TiO_x$[13,17,18], $MgO_x$[19,20] and $ZnO$[21,22], can be used as the electron selective transport layers instead of n-type a-Si:H in c-Si solar cells.

In recent years, remarkable achievements have been made on the SHJ solar cells using molybdenum oxide ($MoO_x$) with wide band-gap (3.0-3.3 eV) and high work function (>6 eV) as the HSL[5,23-27]. Various techniques, such as thermal evaporation[3-5], electron beam evaporation[20,28], atomic layer deposition[29,30], sputtering[31,32] and solution-processed method[6], etc. , are available in preparation of $MoO_x$ thin films. By using thermal evaporated $MoO_x$ thin films as the HSL, silicon heterojunction solar cells with the power conversion efficiencies up to 22.5% were fabricated by Jonas Geissbuhler et al.[3]. Jing Yu et al.[20] prepared the $MoO_x$ films using electron beam evaporation and achieved an efficiency of 14.2% on a plane SHJ solar cell with the structure of $MoO_x$/n-type c-Si/MgO. A solution-processed method was reported by Jingnan Tong et al.[6] to form the $MoO_x$ layers by spin-coating hydrogen molybdenum bronze solution on crystalline silicon wafer surfaces. However, properties of the $MoO_x$ thin films and performance of the devices need to be further improved. To promote the commercialization of the novel SHJ solar cells in the future, a simple and scalable production technique capable of fabricating high quality $MoO_x$ thin films is needed.

In this work, a hot wire oxidation-sublimation deposition (HWOSD) technique was developed to fabricate amorphous molybdenum oxide thin films with good photoelectric properties. SHJ solar cells were fabricated making use of the HWOSD $MoO_x$ as the HSL. Investigations and optimizations of the device structure, interface passivation and annealing process were carried out. A power conversion efficiency up to 21.10% was achieved for a champion SHJ solar cell with the structure of Ag/ITO/n-type a-Si:H/ intrinsic a-Si:H/n-type textured c-Si/intrinsic a-Si:H/MoO$_x$/Ag. Dark *J-V-T* characteristics were analysed to understand the transport mechanisms of the novel heterojunction solar cells.

2. **The hot wire oxidation-sublimation deposition technique**

A schematic diagram of the hot wire oxidation-sublimation deposition (HWOSD) technique is presented in Fig. 1. In a deposition chamber with oxygen atmosphere, molybdenum wires is electrically heated to high temperature. MoO$_x$ molecules are generated on the surface of hot molybdenum wires and are sublimated directly into the chamber. The MoO$_x$ molecules adsorb, diffuse, coalescence and finally form the MoO$_x$ thin films on the substrate.

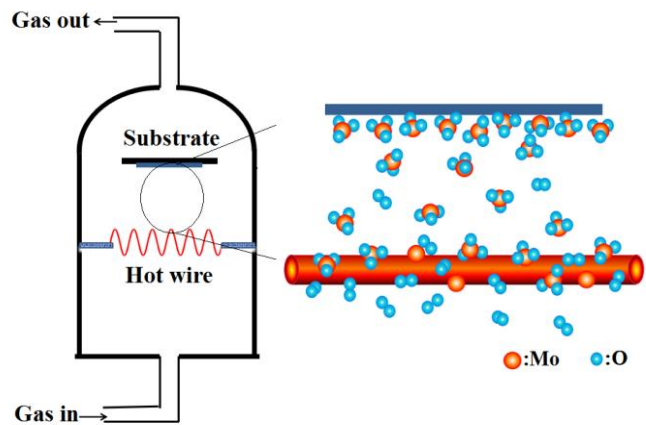

**Fig.1** Schematic diagram of hot wire oxidation-sublimation deposition technique.

The HWOSD technique is novel and it has many advantages over thermal evaporation[3-5] in the preparatin of MoO$_x$ films. For example, by designing the structure of hot wires, for example increasing the length of molybdenum wire and arranging the multiple molybdenum wires, MoO$_x$ films with large area can be deposited by HWOSD; The MoO$_x$ is generated while sublimating in HWOSD, the deposition rate can be increased by raising oxygen pressure and/or the temperature of molybdenum wire without the problem of the powder spattering in the thermal evaporation. Its characteristics of large area and high speed are beneficial to the industrialization. Moreover, the HWOSD technique is based on high melting point of molybdenum and low boiling point of Molybdenum oxide. Therefore, it is suitable to other materials with similar characteristics, such as $WO_x$ and $VO_x$.

## 3. Results and discussion
3.1 Characteristics of the $MoO_x$ thin films prepared by HWOSD

During the process of hot wire oxidation-sublimation deposition, hot wire temperature and oxygen pressure are two important parameters affecting the opto-electronic properties of the deposited $MoO_x$ thin films. Under the optimized hot wire temperature (1095±5℃) and oxygen pressure (0.2Pa), an average transmittance of 94.2% in the wavelength range of 400-1100 nm can be obtained on a 15nm $MoO_x$ thin film. (The effect of the oxygen pressure on the transmittance of the $MoO_x$ thin films is given in Figure S1). The dark conductivity of the optimized $MoO_x$ film is $1.6 \times 10^{-6}$ S/cm. The XRD spectrum shows that the HWOSD $MoO_x$ thin film is in an amorphous structure (Figure S2).

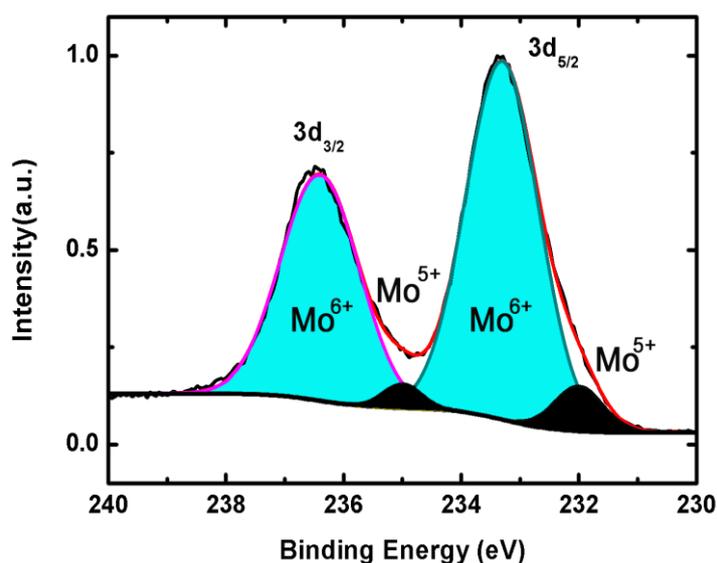

**Fig. 2** The Mo 3d core level XPS spectrum for a $MoO_x$ film fitted with multiple Voigt peaks (shaded areas) to quantify the contribution of different oxidation states.

Figure 2 shows the XPS spectrum of the Mo 3d core level in $MoO_x$ thin film. The two major peaks, with binding energies of 233.3 eV and 236.4 eV, correspond to the $Mo^{6+}$ $3d_{5/2}$ and $3d_{3/2}$, respectively[6,26]. And the minor ones centered at 232.0 eV and

235.0 eV can be attributed to $Mo^{5+}$. The estimated O/Mo atomic ratio of the $MoO_x$ thin film is about 2.94, which closes to the stoichiometric ratio of 3. The relatively high oxygen content in the $MoO_x$ thin films usually leads to a high work function[25,26].

3.2 Surface morphologies and passivation effect of $MoO_x$ on Si substrates

SEM images of the $MoO_x$ thin films deposited on Si wafers are shown in Fig. 3. Figure 3 (a) and (b) exhibit the top-view and cross-sectional SEM images of $MoO_x$ films on polished Si substrate, respectively. It can be seem that compact $MoO_x$ films with high thickness uniformity were formed on the polished Si substrates by using HWOSD. The SEM images of pyramid-shaped silicon surfaces covered with HWOSD $MoO_x$ thin films are shown in Fig. 3 (c) and (d). Conformal coverage of the $MoO_x$ thin films on the random pyramids is realized. The nice coverage characteristics can be confirmed by the elemental EDS mappings of O and Mo (Figure S3).

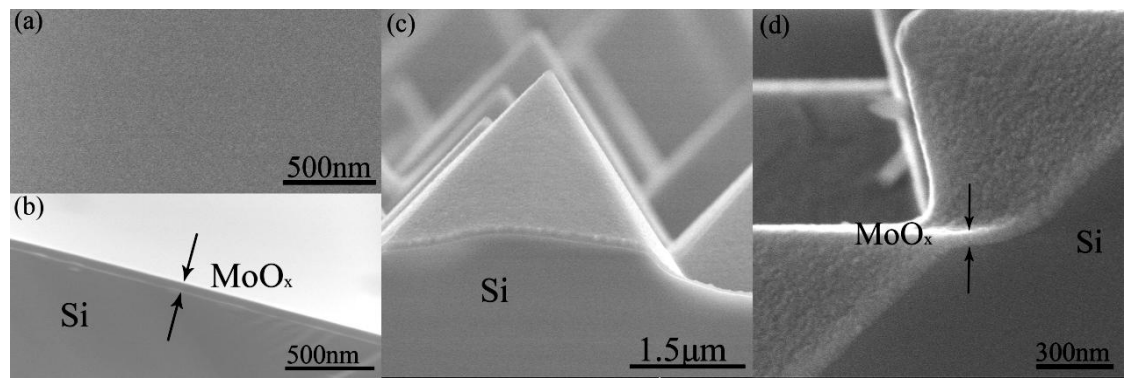

Fig. 3 (a) Top-view SEM image of $MoO_x$ (25nm) on polished Si substrate. (b) Cross-sectional SEM image of $MoO_x$ (70 nm) on polished Si substrate. (c, d) SEM images of $MoO_x$ (70 nm) on textured Si wafers.

Figure 4 shows the measured effective lifetimes for n-type CZ Si wafers sandwiched between i a-Si:H thin film or i a-Si:H/$MoO_x$ combination layer as a function of excess carrier density. Compared with the effective lifetimes of the a-Si:H passivated c-Si wafer, the a-Si:H/$MoO_x$ combination layer passivated sample shows enhanced lifetimes in the whole carrier injection concentration range of $7 \times 10^{14}$ $cm^{-3} \sim 1 \times 10^{16} cm^{-3}$ and a maximum lifetime up to 1.3ms is achieved. The results demonstrate the effective field passivation effect of the molybdenum oxide film on the surface of c-Si. The high work function of the $MoO_x$ layer produces a large band bending of the n-type c-Si surface and thus reduces the surface recombination effectively.

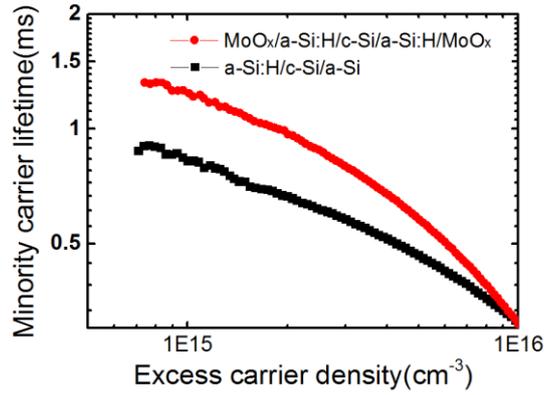

**Fig. 4** Measured effective lifetimes for samples with the structures of a-Si:H/c-Si/a-Si:H and MoO$_x$/a-Si:H/c-Si/a-Si:H/MoO$_x$ as a function of excess carrier density in the range of $7\times10^{14}$ cm$^{-3}$~$1\times10^{16}$ cm$^{-3}$.

3.3 Optimization of the SHJ solar cells with MoO$_x$ as HSL

The structure of the heterojunction solar cells with MoO$_x$ HSL is shown as Fig. 5(a). The illuminated side (front side) structure of the devices is set to be Ag grid/ITO/n-type a-Si:H/ intrinsic a-Si:H. The rear side structure is Ag/MoO$_x$/intrinsic a-Si:H.

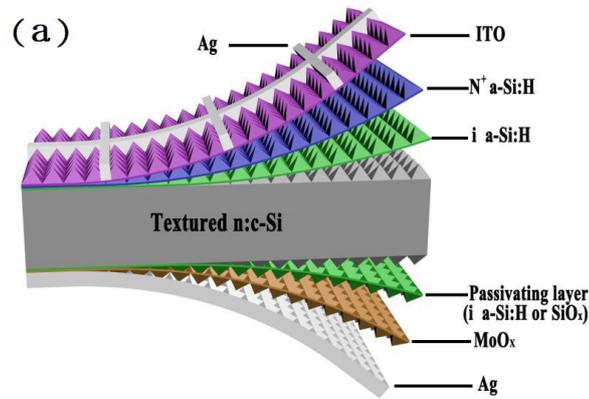

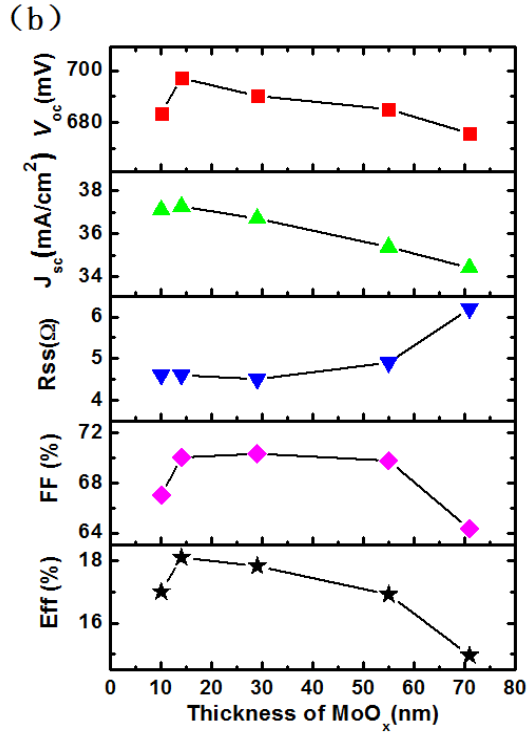

**Fig. 5** (a) Cross-sectional schematic of the heterojunction solar cells with MoO$_x$ HSL. (b) Photovoltaic parameters of the SHJ solar cells varied with different thicknesses of MoO$_x$ HSL.

Effective interface passivation plays critical role in increasing the performance of SHJ solar cells. We found that SHJ solar cells with MoO$_x$ HSL directly deposited on Si substrates usually show much worse performance than a traditional SHJ solar cell, implying a serious recombination of photo-generated carriers at the MoO$_x$/c-Si interface. Different passivation layers, including intrinsic a-Si:H thin film, UV/O3 photo-oxidized SiO$_x$, and the combination of a-Si:H and SiO$_x$, were adopted to passivate the c-Si substrates before the deposition of MoO$_x$ (Table S1). Compared with the solar cell without any passivation layer, all the passivation methods we tried improved the performances of the solar cells to some extent. Among them, the intrinsic a-Si:H layer deposited using PECVD exhibits the best passivation effect. The UV/O3 treatment is quite simple compared with the a-Si:H deposition technique. However, the passivation effect of the UV/O3 photo-oxidized SiO$_x$ layer is less than satisfactory. Similar to reported in the literature[3], we also noticed that the SiO$_x$ layer may lead to the deterioration of the annealing performance of a SHJ solar cell with MoO$_x$ HSL. Therefore, we eliminated the UV/O3 photo-oxidized SiO$_x$ and used the a-Si:H as the passivation layer at present stage.

Using intrinsic a-Si:H as the interface passivation layer, influence of the MoO$_x$

HSL thickness on the photovoltaic parameters of the SHJ solar cells was investigated as shown in Fig. 5(b). Increasing the $MoO_x$ HSL thickness from 10 nm to 14 nm raises the $V_{oc}$ of the solar cells by 13.5 mV (from 683.6mV to 697.1mV). However, the $V_{oc}$ goes down instead of rising as we further increased the $MoO_x$ HSL thickness. And, it drops to 675.8mV at a $MoO_x$ thickness of 71 nm. In the case that the thickness of the $MoO_x$ HSL is too small, the space charge region near the c-Si surface may not be well formed, which is not favorable for the separation of the photo-generated carriers. However, as for the condition of excessive thickness, the consequentially enhanced recombination in the $MoO_x$ layer leads to the decline of the open circuit voltage. At the same time, the reduced carrier collection causes a rapid drop in the short circuit current density ($J_{sc}$) as Fig. 5(b) shows.

Fill factor (FF) is affected simultaneously by energy band structure and series resistance. The small FF in the 10nm $MoO_x$ condition could also be a consequence of the insufficient build-up of the space charge region. When the thickness of $MoO_x$ is between 14nm and 55nm, the relatively large built-in voltage and the almost invariable series resistance keep the FF at a relatively high value. When the thickness of the $MoO_x$ reaches 70nm, the high series resistance results in a significant reduction in FF. Considering all the photovoltaic parameters of the series of SHJ solar cells, we determined an optimal $MoO_x$ thickness of 14nm. In this case, $MoO_x$ with high work function and sufficient thickness promotes the establishment of the space charge region near the c-Si surface, and extra carrier recombination related to excessive $MoO_x$ thickness can be avoided. A power conversion efficiency of 18.11% was achieved for the SHJ solar cell with a 14nm $MoO_x$ as the HSL.

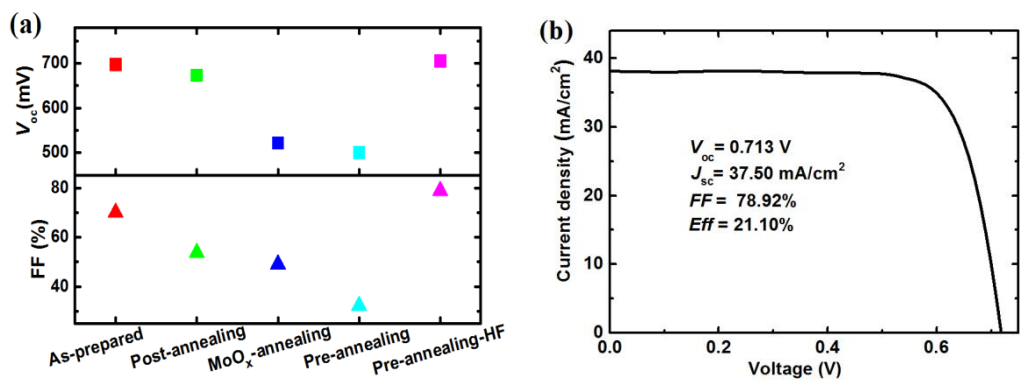

**Fig.6** (a) Influence of annealing process on $V_{oc}$ and FF of the SHJ solar cells with $MoO_x$ HSL.
(b) Light J-V characteristic of a champion SHJ solar cell with $MoO_x$ HSL.

Post-annealing is a common process during the fabrication of conventional SHJ

solar cells, which can improve the overall performance of the devices. However, post-annealing tends to show a bad effect on the SHJ solar cells with metal oxides (such as $MoO_x$, $WO_x$, $VO_x$) as the hole selective transport layer[3,7]. Possible reasons are that the work functions of metal oxides and the characteristics of the interface between the metal oxides and the c-Si have been changed by post-annealing [33].

In this paper, a preliminary investigation on the annealing process of the SHJ solar cells with $MoO_x$ HSL was carried out and the influences of different annealing processes on the $V_{oc}$ and FF of the SHJ solar cells are illustrated in Fig. 6 (a). As a reference, the $V_{oc}$ and FF of an as-prepared SHJ solar cell without undergoing any annealing process are also included as shown in red symbols. Similar to reported in the literature, annealing after the fabrication of the whole device (post-annealing, 190 °C, 5min) leads to deterioration in device performance (green symbols). $MoO_x$-annealing (blue symbols) in Fig. 6 (a) refers to an annealing process that was carried out just after the preparation of $MoO_x$ and before the deposition of silver electrode. $V_{oc}$ and FF of the solar cell are further reduced with $MoO_x$-annealing compared with the post-annealing process, indicating serious damage to the exposed $MoO_x$ layer was caused by the $MoO_x$-annealing process. In order to avoid the annealing damage to the $MoO_x$ layer, annealing process was carried out before the preparation of the $MoO_x$ layer, namely pre-annealing. The solar cell performance, disappointingly, is worse instead of getting better. The possible reason is that an a-Si:H/$SiO_x$ double-layer is formed during the pre-annealing process in the air atmosphere. Therefore, we removed the oxidation layer on the amorphous silicon by HF solution (2%) treatment after the pre-annealing process and immediately prepared the $MoO_x$ HSL and the metal back electrode to finish the device fabrication. Just as expected, the performance of the pre-annealing-HF sample shows a remarkable improvement as depicted in Fig. 6(a). Figure 6 (b) is the light J-V curve of an optimal solar cell fabricated with the pre-annealing-HF process. An efficiency of 21.10% with $V_{oc}$ of 713 mV, $J_{sc}$ of 37.50 $mA/cm^2$ and FF of 78.92% was achieved for the champion SHJ solar cell with a $MoO_x$ HSL.

3.4 Transport mechanisms of the SHJ solar cells with $MoO_x$ HSL

To further improve the performance of the SHJ devices, it is essential to understand the mechanisms governing charge carrier transport of the novel heterojunction. It is well known that an analysis of the dark *J-V-T* characteristics can provide a deep understanding of the transport mechanisms. Figure 7 (a) shows the dark *J-V* curves of a SHJ solar cell with the structure of Ag/ITO /n$^+$ a-Si:H /i a-Si:H/n c-Si/i a-Si:H/ MoO$_x$/Ag measured at temperatures from 200K to 380K. The corresponding photovoltaic parameters of the device are V$_{oc}$=710mV, FF=77.0%, J$_{sc}$=37.3mA/cm$^2$ and Eff=20.4%.

According to the universal rectification models, the relation between the current density and the applied voltage can be written as an empirical equation:

$$J = J_0[e^{AV} -1] \quad (1)$$

where $J_0$ is the saturation current density and the exponential factor *A* depends on the transport mechanisms.

Based on equation (1), the *J-V* curves were fitted in a lower voltage range (0.1 - 0.4V), in which the transport mechanisms can be better investigated. By studying the temperature dependences of the fitting parameters $J_0$ and *A* in equation (1) (Figure S4), we found that an approximate linear relationship between $ln(J_0)$ and $1/kT$ can be determined. However, *A* does not change significantly in the whole temperature range and the *A* and $1/kT$ relationship seriously deviates from linearity in both high and low temperature regions. According to the empirical expression *A=q/nkT*, where *q* is the electron charge, *n* the diode ideality factor, *k* the Boltzmann's constant, *T* the absolute temperature, no unified ideality factor *n* can be determined in the entire temperature range. For a relatively narrower temperature region of 250*K<T<*330*K* (35.0<1/*kT*<46.2), fairly large diode ideality factor values (3.51~3.93) can be obtained.

In the traditional SHJ solar cells, large *n* is usually considered to be a signal of serious interface recombination which certainly degrades the V$_{oc}$[34]. In this paper, the contradiction between the large *n* and quite high V$_{oc}$ (710mV) of the novel SHJ device implies different transport mechanisms from the traditional ones. Other transport channels, like tunnel, probably play important roles in the carrier transportation of the novel SHJ solar cells.

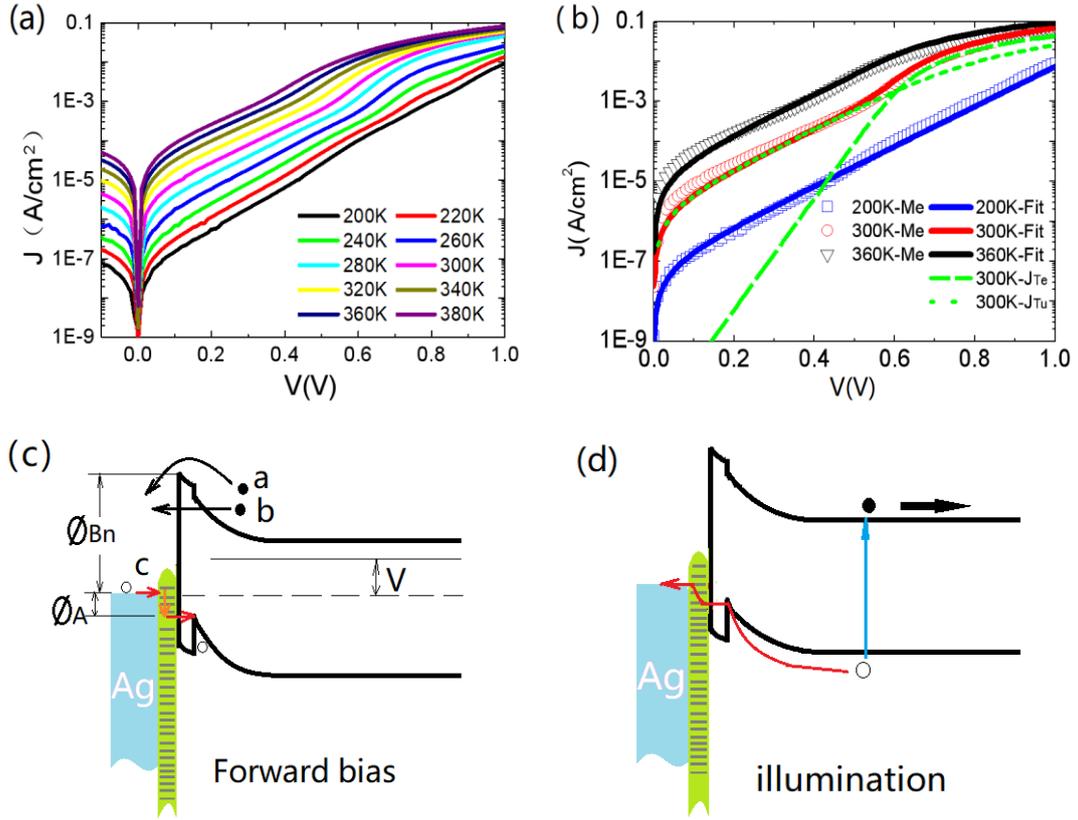

Figure 7 (a) Dark *J-V* curves of a novel SHJ solar cell measured at the temperatures from 200K to 380K. (b) Experimental (hollow symbols) and fitting ($J=J_{Te}+J_{Tu}$, solid lines) dark *J-V* curves under forward bias for the novel SHJ solar cell under three temperatures of 200K, 300K and 360K. The green dashed and dotted lines respectively represent the fitting $J_{Te}$ (majority-carrier processes a and b) and $J_{Tu}$ (minority-carrier process c) for the *J-V* curve measured under 300K. (c) Schematic energy band structures of the MoOx/i a-Si:H/n c-Si heterocontact under forward bias. Three transport processes, including thermionic emission of electrons, thermal assisted tunneling of electrons and tunneling of holes are denoted by a, b, and c, respectively. (d) Schematic energy band structures of the heterocontact under illumination.

Considering the high work function (>6eV) and high density of gap states characteristics of amorphous $MoO_x$[5], it is more appropriate to think of the $MoO_x$/i a-Si:H/n c-Si heterocontact as a Schottky junction, whose energy band structures under forward bias and illumination are depicted in Fig. 7 (c and d). Three main transport processes, including thermionic emission of electrons, thermal assisted tunneling of electrons and quantum assisted tunneling of holes are denoted by a, b, and c, respectively. In both a and b processes, thermionic activation of majority carriers is involved. Therefore, we use a total majority-carrier (electron) thermionic current density $J_{Te}$ to represent the thermionic emission and thermal assisted tunneling processes. Considering the series resistance $R_s$ at the same time, $J_{Te}$ can be expressed as

$$J_{Te} = J_{S1}\{\exp[\frac{q(V-JR_s)}{nkT}] - 1\} \qquad (2)$$

$$J_{S1} = C(T)\exp(-\frac{q\phi_{Bn}}{kT})$$

where $\phi_{Bn}$ is the effective barrier height for electrons, C(T) is a temperature dependent pre-factor.

The process c in Fig. 7(c) presents the tunneling process of holes. The current density $J_{Tu}$ can be expressed as [35,36]

$$J_{Tu} = J_{S2}\{\exp[\frac{q(V-JR_s)}{E_0}] - 1\} \qquad (3)$$

$$J_{S2} = B(T)\exp(-\frac{q\phi_A}{kT})$$

In this expression, $E_0$ is named as tunneling barrier energy and it corresponds to the almost invariable slopes of the dark J-V curves at low bias as Fig. 7(a) shows. In the following fittings, $E_0$ is simplified to be a constant. $q\phi_A$ is the difference between the top of the valence band (TVB) of c-Si surface and the Fermi lever of Ag electrode, as shown in Fig. 7(c). Holes tunnel through the gap states of the $MoO_x$ layer into the TVB of c-Si and then recombine with electrons. It is actually a minority-carrier behavior.

The total current density $J$ is given by the sum of $J_{Te}$ and $J_{Tu}$. Three forward J-V curves measured at temperatures of 200K, 300K, and 360K were fitted according to $J = J_{Te} + J_{Tu}$, as shown in Fig. 7(b). Except for the very low voltage region, basic agreement between the measured and the fitted curves in the wide voltage range can be obtained. At this time, a more reasonable ideality factor of n=1.2, which is matched with the high $V_{oc}$, is obtained. And the fitting results for $E_0$, $q\phi_A$, and $q\phi_{Bn}$ are 0.087eV, 0.31eV, and 0.79eV, respectively. The sum of the fitted $q\phi_A$ and $q\phi_{Bn}$ is 1.1eV, which coordinates with the band gap of c-Si. The large $q\phi_{Bn}$ indicates a strong band bending at the n-type c-Si surface, which makes the surface inversion and promotes the collection of photo-generated electrons, as shown in Fig 7(d). The slight deviation between the fitting curves and the experimental ones in the very low voltage region is probably related to the contributions from shunt resistance and recombination processes[34]. To better understand the respective contributions of $J_{Tu}$ and $J_{Te}$ on the total *J-V* curve, the fitted $J_{Tu}$ and $J_{Te}$ for the *J-V* curve measured under 300K, as an example, are depicted in dashed and dotted lines in Fig. 7 (b) . It can be seem that $J_{Te}$ starts to play a major role as *V*>0.6V. In the case of lower forward bias (V<0.5V), the tunnel current density $J_{Tu}$ dominates the *J-V* curve and causes a lower slope of the plot of *Log*(*J*) versus *V*. The high defect density in the band gap of amorphous $MoO_x$ makes the tunneling process easily and the minority current density $J_{Tu}$ is larger than that of a typical metal/Si Schottky junction. This tunneling channel is critical important for the carriers collection under illumination. Photo-generated holes can be effectively collected through this tunneling channel as shown in Fig. 7 (d), which is beneficial to obtain high open-circuit voltage.

## 4. Conclusions

The hot wire oxidation-sublimation deposition method is demonstrated to be a promising technique for preparation of high-quality $MoO_x$ films with uniform thickness, compact structure, nice photoelectric properties and conformal coverage on textured Si substrates. Novel silicon heterojunction solar cells with the HWOSD $MoO_x$ thin films as the HSL were successfully fabricated. Finally, an efficiency of 21.10% was achieved for the champion SHJ solar cell with a $MoO_x$ HSL fabricated by the scalable HWOSD technique.

Analysis of the dark *J-V-T* characteristics shows that tunneling of holes through the gap states of the $MoO_x$ layer causes the low slope of the plot of *Log* (*J*) versus *V* at low voltage range (0.1*V* - 0.4*V*). Different from the traditional SHJ solar cells, the low *Log*(*J*)-*V* slope or high ideality factor *n* is no longer a sign of low $V_{oc}$ for the novel SHJ solar cell. $V_{oc}$ as high as 710 mV can be achieved for the novel SHJ solar cell. The tunneling channel in valence band plays an important role in holes collection under illumination.

Besides the advantages of reducing light absorption loss and lowing HSL fabrication costs, the nature of large build-in voltage (represented by high $\phi_{Bn}$) and effective carrier collection makes the $MoO_x$/c-Si heterojunction inherently possess high efficiency potential. More research work is needed to reveal the potential of the novel SHJ solar cells.

## 5. Experimental details

5.1 Preparation of MoO$_x$ thin films by HWOSD

During the deposition, the substrate temperature was lower than 70℃. The molybdenum wires with a purity of 99.995% and a diameter of 1 mm were used. The hot wire temperature, oxygen flow rate and deposition pressure were optimized to be 1095±5℃, 4 sccm and 0.2 Pa, respectively.

5.2 Fabrication of silicon heterojunction solar cells

N-type <100> float zone (FZ) silicon wafers with a thickness of 250 μm and a resistivity of 1 to 5 Ω·cm were used as the substrates. Alkaline texturing and isotropic etching were carried out in succession to form random pyramids with less sharp tops on both sides of the c-Si wafers. After texturing, the c-Si substrates were chemically cleaned according to the standard RCA procedure. Before proceeding to the next step, the textured silicon wafers were dipped in 2% hydrofluoric acid for 1 min to remove the surface oxide layer.

The illuminated side (front side) structure of the devices is set to be Ag grid/ITO/n-type a-Si:H/ intrinsic a-Si:H. Intrinsic (~7 nm) and n-type (~10 nm) a-Si:H as the passivation and the electron selective transport layer were successively deposited on one side of the c-Si substrates by means of plasma enhanced chemical vapor deposition (PECVD) under a substrate temperature of about 200 °C. The front electrode consists of an indium tin oxide layer (ITO 80nm) by magnetron sputtering and a Ag grid by thermal evaporation. The rear side structure of the SHJ solar cells is Ag/MoO$_x$/intrinsic a-Si:H. The intrinsic a-Si:H (i a-Si:H) thin film (~6nm) deposited by PECVD was also used for interface passivation. The MoO$_x$ thin film, as the HSL to replace the traditional p-type a-Si:H, was prepared by hot wire oxidation-sublimation deposition. Thermal evaporated Ag film was used as the rear electrode. The active area of the devices is 1 cm$^2$.

5.3 Characterization of thin films and devices

Thicknesses of MoO$_x$ and ITO thin films were measured using a surface profilometer (ERUKER-DektakXT). Surface morphologies and elemental analysis of the MoO$_x$ thin films deposited on c-Si wafers were characterized using scanning electron microscope (SEM, Hitachi SU8010) and energy dispersive spectrometer (EDS, Bruker 6-30). Elemental compositions of the MoO$_x$ films were characterized using X-ray photoelectron spectroscopy (XPS, Thermo Scientific ESCALAB250Xi) under ultra-high vacuum (<2×10$^{-9}$ mbar). Minority carrier lifetimes of the passivated c-Si wafers were evaluated using the quasi-steady-state photo conductance (QSSPC Sinton WCT-120). Light current density-voltage (*J-V*) curves of the solar cells were obtained under AM1.5 (100 mW/cm$^2$, 25℃) illumination. Dark *J-V-T* characteristics of the SHJ solar cells were measured at the temperatures from 200K to 380K.

## Acknowledgements

This work was supported by the National Natural Science Foundation of China (No.61604153 and No.61674150). Support from the Sino-Danish Center for Education and Research (SDC) is fully acknowledged.